\documentclass{jltp}
\usepackage{graphicx}
\begin{document}

\title{Magnetoresistance of UPt$_3$}

\author{T.M. Lippman, J.P. Davis, H. Choi, J. Pollanen,\\ and W.P. Halperin}

\address{Department of Physics and Astronomy, Northwestern University,\\ Evanston, IL 60208, USA}

\runninghead{T.M. Lippman \textit{et al.}}{Magnetoresistance of UPt$_3$}

\maketitle

\begin{abstract}
We have performed measurements of the temperature dependence of the magnetoresistance up to 9 T in bulk single crystals of UPt$_3$ with the magnetic field along the \textit{b-}axis, the easy magnetization axis. We have confirmed previous results for transverse magnetoresistance with the current along the \textit{c-}axis, and report measurements of the longitudinal magnetoresistance with the current along the \textit{b-}axis. The presence of a linear term in both cases indicates broken orientational symmetry associated with magnetic order. With the current along the \textit{c-}axis the linear term appears near 5 K, increasing rapidly with decreasing temperature. For current along the \textit{b-}axis the linear contribution is negative.

\end{abstract}

\section{INTRODUCTION}
The heavy fermion system UPt$_3$ exhibits unconventional superconductivity coexisting with magnetic order\cite{joynttaillefer02}. The superconducting transition occurs at $T_c$ = 563 mK in the clean limit\cite{kycia98}, while the antiferromagnetic transition occurs at $T_N\approx$ 5 K. The antiferromagnetic transition was first observed\cite{heffner89} in $\mu$SR and then confirmed by neutron scattering\cite{aeppli88}. However, a subsequent $\mu$SR study on a different sample showed no effect\cite{dalmas95}, and the transition remains unseen in thermodynamic measurements\cite{fisher91} or nuclear magnetic resonance\cite{tou96}. In addition, a single report by Behnia \textit{et al.}\cite{behnia90} of the onset of a linear term in the magnetoresistance at 5 K gives another indication of antiferromagnetic ordering. To explain these combined results it has been suggested that the magnetic peak observed at 5 K in neutron scattering is due to antiferromagnetic fluctuations, rather than static order, while the true magnetic transition is at a lower temperature\cite{lee93,flouquet05}. This is identified with anomalies seen in the heat capacity\cite{schuberth92,brison94} at 18 mK as well as in the magnetization\cite{schottl99}. Our knowledge of UPt$_3$ would be greatly improved by a full understanding of its magnetic structure, from 10 K down to 10 mK. Consequently, we have revisited the question of magnetoresistance with high quality bulk samples, since magnetoresistance appears to be the only measurement other than neutron scattering that consistently indicates the existence of a magnetic transition near 5 K.


\section{EXPERIMENTAL DETAILS}
\begin{figure}[t]
\begin{center}
		\includegraphics[width=0.65\textwidth]{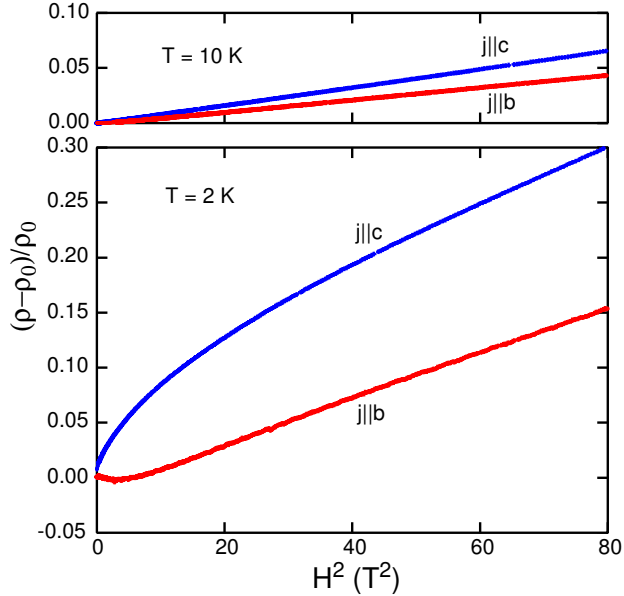}
	\caption{\label{compare}(Color online) Magnetoresistance of UPt$_3$ at 2 and 10 K, vs. H$^2$ with magnetic field H$\left|\right|$b. At high temperatures the magnetoresistance is a pure quadratic, as expected from a simple symmetry argument. Linear contributions to the magnetoresistance are evident at low temperature.}
\end{center}
\end{figure}
Our samples are cut from a single crystal of UPt$_3$ grown by electron beam vertical float zone refining under ultrahigh vacuum (UHV). The crystal axes are then determined by Laue x-ray scattering and needles cut out by electro-discharge machining followed by etching in aqua regia. One sample was cut with its length along the \textit{c-}axis and annealed in a UHV electron bombardment furnace at 800 $^{\circ}$C for six days. It has residual resistance ratio RRR$_c$ = 890, superconducting transition temperature $T_c$ = 551.5 mK, and superconducting transition width $\delta T_c$ = 5.7 mK. The second sample was cut with its length along the \textit{b-}axis and annealed at 970 $^{\circ}$C for six days. It has residual resistance ratio RRR$_c$ = 957, superconducting transition temperature $T_c$ = 549.0 mK, and superconducting transition width $\delta T_c$ = 2.3 mK. The samples were then secured to a microscope slide using Stycast 2850 epoxy. Electrical connections were made using Cu wires attached with Pb-Sn solder. The samples were then placed in a gas flow cryostat and held at constant temperature while measuring the resistance from 0 to 9 Tesla using an LR-700 AC resistance bridge. Here we present the results of this procedure with the current along both the \textit{b-} and \textit{c-}axes and the field perpendicular to the \textit{c-}axis.

\section{RESULTS}
\begin{figure}[t]
\begin{center}
		\includegraphics[width=0.75\textwidth]{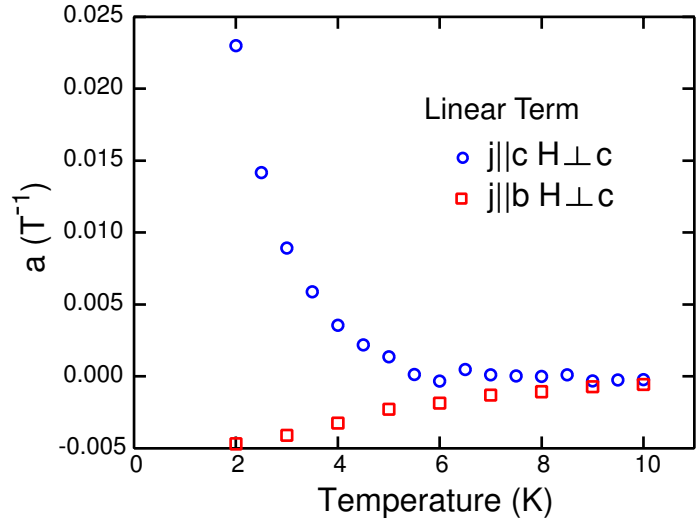}
	\caption{\label{avsT}(Color online) The linear contribution to the magnetoresistance. Note the strong increase at 5.5 K with the current along the \textit{c-}axis, associated with the antiferromagnetic transition. A weaker and negative linear magnetoresistance is observed with the current along the \textit{b-}axis}
\end{center}
\end{figure}

\begin{figure}[t]
\begin{center}
		\includegraphics[width=0.75\textwidth]{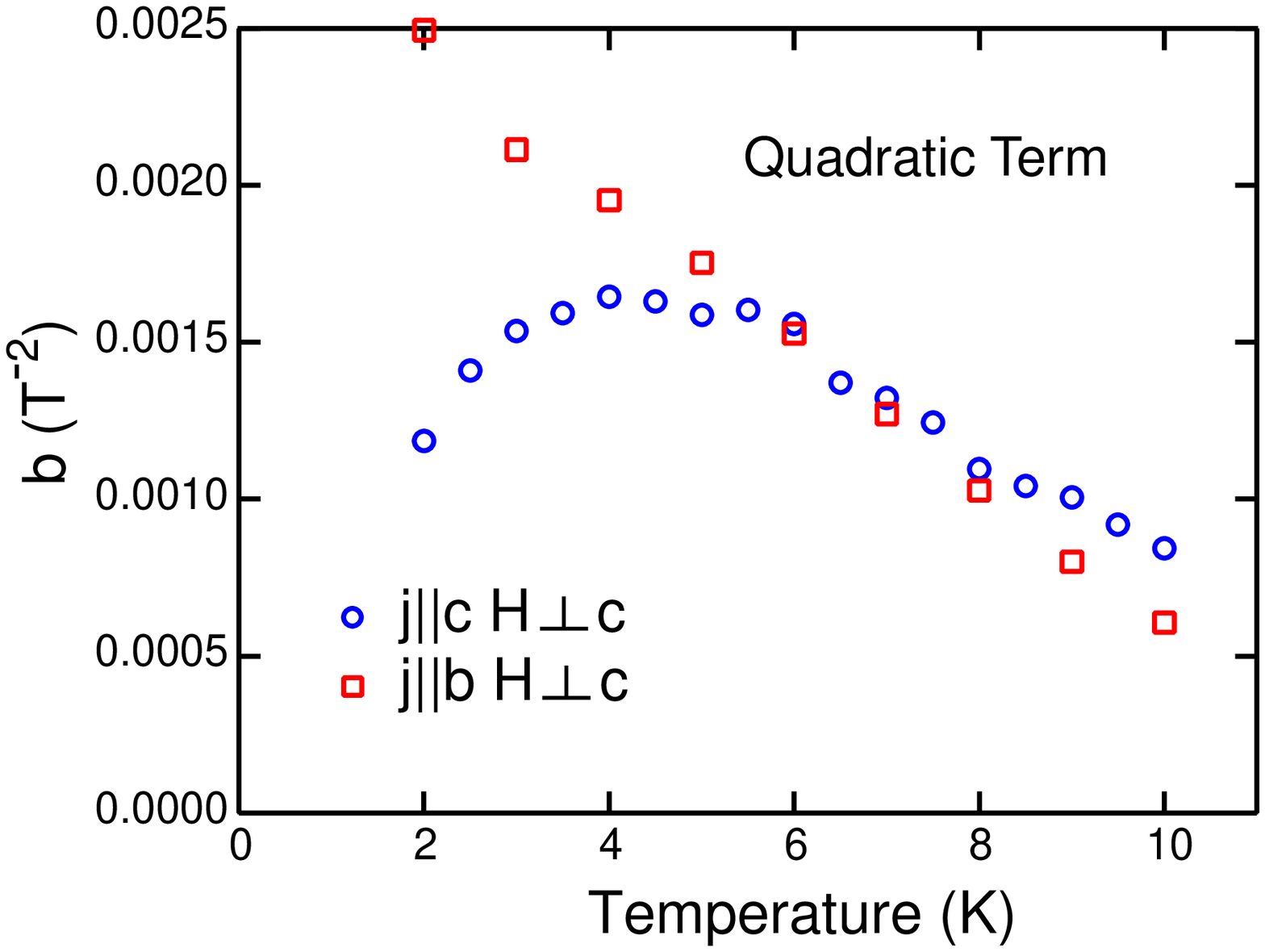}
	\caption{\label{quad}(Color online) The quadratic term of the fit procedure. The two samples have the same behavior above 6 K, when the \textit{c-}axis data breaks off due to an antiferromagnetic transition.}
\end{center}
\end{figure}

Data from 2 and 10 K are shown in Fig.~\ref{compare}, plotted against the square of the magnetic field. The data at all temperatures were found to be well fit by a polynomial of the form:
\begin{equation}
	\frac{\rho-\rho_{0}}{\rho_{0}} = aH + bH^{2},
\end{equation}
where H is the applied field in Tesla, $\rho_{0}$ is the resistivity in zero field, and the fit coefficients $a$ and $b$ are functions of temperature, shown in Figs.~\ref{avsT} and ~\ref{quad} respectively. 
For the \textit{c-}axis sample and transverse field, the linear term of the fit is zero from 10 K down to 5.5 K. Below 5 K, the linear term increases rapidly, with no sign of saturation. The quadratic term increases as temperature is decreased, with a maximum near 4 K. For the \textit{b-}axis sample and longitudinal field, the linear term has opposite sign and smaller magnitude compared with the linear term in the \textit{c-}axis sample. Also, it becomes non-zero closer to 10 K, and does not change slope with temperature as dramatically as the \textit{c-}axis data. The quadratic term similarly shows no change in slope for the entire temperature region studied. The quadratic term for both samples increases with decreasing temperature above 6 K, below which the \textit{c-}axis data breaks away from the \textit{b-}axis data and decreases.

Behnia \textit{et al.}\cite{behnia90} have previously measured the transverse magnetoresistance of a monocrystalline whisker with current along the \textit{c-}axis. Our results using a bulk single crystal with current along the \textit{c-}axis confirm their results. However, slight differences in the behavior of the fit coefficients are due to the fact that our data were first normalized by dividing by the zero field resistivity. That is, Behnia \textit{et al.} fit to $\rho-\rho_{0}$, while we fit to $(\rho-\rho_{0})/\rho_{0}$. The primary effect of this is to move the maximum in the \textit{c-}axis quadratic term from 8 K down to 4 K. Comparing longitudinal and transverse magnetoresistance we see that the temperature where they begin to deviate from one another is approximately 6 K for both the linear and the quadratic contributions.

\section{CONCLUSIONS}
The presence of a linear term in the magnetoresistance is a sign of broken inversion symmetry, implying that the system has a preferred orientation which we associate with magnetic order. With the current along the \textit{c-}axis and the field transverse, this occurs near 6 K. We associate the onset and subsequent rapid increase of the linear term as temperature decreases with the antiferromagnetic Bragg peak observed by neutron scattering\cite{aeppli88}. The absence of similar behavior with the current along the \textit{b-}axis, field longitudinal, demonstrates anisotropy in the magnetic order. We speculate that the negative linear magnetoresistance is an indication of suppression of a channel for spin-flip scattering that grows with increasing antiferromagnetic order parameter. Within the resolution of our measurement, we do not observe a discontinuity in temperature dependence which might be associated with a thermodynamic phase transition.

\section*{ACKNOWLEDGMENTS}
We would like to thank O. Chernyashevskyy for technical assistance. We acknowledge support from the Department of Energy, DOE grant DE-FG02-05ER46248, and the Weinberg College of Arts \& Sciences.


\begin{thebibliography}{99}
\bibitem{joynttaillefer02}
R. Joynt and L. Taillefer, Rev. Mod. Phys. \textbf{74}, 235, (2002).

\bibitem{kycia98}
J.B. Kycia, J.I. Hong, M.J. Graf, J.A. Sauls, D.N. Seidman, and W.P. Halperin, Phys. Rev. B \textbf{58}, R603, (1998).

\bibitem{heffner89}
R.H. Heffner, D.W. Cooke, A.L. Giorgi, R.L. Hutson, M.E. Schillaci, H.D. Rempp, J.L. Smith, J.O. Willis, D.E. MacLaughlin, C. Boekema, R.L. Lichti, J. Oostens, and A.B. Denison, Phys. Rev. B \textbf{39}, 11345, (1989).

\bibitem{aeppli88}
G. Aeppli, E. Bucher, C. Broholm, J.K. Kjems, J. Baumann, and J. Hufnagl, Phys. Rev. Lett. \textbf{60}, 615, (1988).

\bibitem{dalmas95}
P. Dalmas de R\'{e}otier, A. Huxley, A. Yaouanc, J. Flouquet, P. Bonville, P. Imbert, P. Pari, P.C.M. Gubbens, and A.M. Mulders, Phys. Lett. A \textbf{205}, 239, (1995).

\bibitem{fisher91}
R.A. Fisher, B.F. Woodfield, S. Kim, N.E. Phillips, L. Taillefer, A.L. Giorgi, and J.L. Smith, Solid State Commun. \textbf{80}, 263, (1991).

\bibitem{tou96}
H. Tou, Y. Kitaoka, K. Asayama, N. Kimura, Y. \={O}nuki, E. Yamamoto, and K. Maezawa, Phys. Rev. Lett. \textbf{77}, 1374, (1996).

\bibitem{behnia90}
K. Behnia, O. Laborde, L. Taillefer, and J. Flouquet, Physica B \textbf{165 \& 166}, 431, (1990).

\bibitem{lee93}
M. Lee, G.F. Moores, Y.-Q. Song, W.P. Halperin, W.W. Kim, and G.R. Stewart, Phys. Rev. B \textbf{48}, 7392, (1993).

\bibitem{flouquet05}
J. Flouquet, Progress in Low Temp. Phys.  \textbf{15}, 139, (2005).

\bibitem{schuberth92}
E.A. Schuberth, B. Strickler, and K. Andres, Phys. Rev. Lett. \textbf{68}, 117, (1992).

\bibitem{brison94}
J.P. Brison , N. Keller, P. Lejay, J.L. Tholenee, A. Huxley, N. Bernhoeft, A.I. Buzdin, B. F\.{a}k, J. Flouquet, L. Schmidt, A. Stepanov, R.A. Fisher, N. Phillips, and C. Vettier, J. Low Temp. Phys. \textbf{95}, 145, (1994).

\bibitem{schottl99}
S. Sch\"ottl, E.A. Schuberth, K. Flachbart, J.B. Kycia, J.I. Hong, D.N. Seidman, W.P. Halperin, J. Hufnagl, and E. Bucher, Phys. Rev. Lett. \textbf{82}, 2378, (1999).

\end{thebibliography}
\end{document}